\documentclass[12pt, letterpaper]{article}

\usepackage{amssymb}
\usepackage{amsmath}
\usepackage{verbatim}

\mathversion{normal}

\makeatletter \@addtoreset{equation}{section} \makeatother

\newcommand{\bea}{\begin{eqnarray}}
\newcommand{\eea}{\end{eqnarray}}
\newcommand{\be}{\begin{eqnarray}}
\newcommand{\ee}{\end{eqnarray}}
\newcommand{\ben}{\begin{eqnarray*}}
\newcommand{\een}{\end{eqnarray*}}
\newcommand{\beq}{\begin{equation}}
\newcommand{\eeq}{\end{equation}}

\newcommand{\id}{\mathbb{I}}
\newcommand{\diag}{\mathop{\rm diag}\nolimits}
\ifx\genfrac\sdflkaj

\else

\fi
\newcommand{\sfrac}[2]{{\textstyle\frac{#1}{#2}}}
\ifx\half\sdflkaj
\newcommand{\half}{\sfrac{1}{2}}
\fi

\let\oldPhi=\Phi
\let\oldPsi=\Psi
\renewcommand{\Phi}{\mathnormal{\oldPhi}}
\renewcommand{\Psi}{\mathnormal{\oldPsi}}

\newenvironment{myeqnarray}{\arraycolsep0pt\begin{eqnarray}}{\end{eqnarray}\ignorespacesafterend}
\newenvironment{myeqnarray*}{\arraycolsep0pt\begin{eqnarray*}}{\end{eqnarray*}\ignorespacesafterend}

\def\[{\begin{equation}}
\def\]{\end{equation}}
\def\<{\begin{myeqnarray}}
\def\>{\end{myeqnarray}}

\ifx\href\asklfhas\newcommand{\href}[2]{#2}\fi

\begin{document}
\thispagestyle{empty}
\begin{flushright}\footnotesize
\texttt{CQUeST-2012-0000}
\end{flushright}
\vspace{.5cm}

\begin{center}%
{\Large\textbf{\mathversion{bold}%
Worldsheet $S$-matrix of $\beta$-deformed SYM}\par}\vspace{1.5cm}%

\textsc{Changrim Ahn$^{\S}$\footnote{ahn@ewha.ac.kr}, Minkyoo Kim$^{\dag}$\footnote{mkim80@sogang.ac.kr}, and Bum-Hoon Lee$^{\dag}$}\footnote{bhl@sogang.ac.kr} \vspace{8mm}

\textit{\small $^\S$ \it Department of Physics and Institute for the Early Universe\\
Ewha Womans University, Seoul 120-750, S. Korea}\\
\vskip .5cm
\textit{\small $^\dag$ \it Department of Physics and Center for Quantum Spacetime\\
 Sogang University, Seoul 121-742, S. Korea}
\vspace{4mm}

\textbf{Abstract}\vspace{7mm}

\begin{minipage}{14.7cm}
We compute perturbative worldsheet $S$-matrix of $\beta$-deformed ${\rm AdS/CFT}$ in the strong and weak `t Hooft coupling limit to compare with exact $S$-matrix.
For the purpose we take near BMN limit of $TsT$-transformed $AdS_5\times S^{5}$ with the twisted boundary condition and compute the $S$-matrix on worldsheet using light-cone gauge fixed Lagrangian.
For the weak coupling side, we compute the $S$-matrix in the $SU(3)$ sector by applying coordinate Bethe ansatz method to the one-loop dilatation operator obtained from the deformed super Yang-Mills theory.
These analysis support the conjectured exact $S$-matrix in the leading order for both sides of $\beta$-deformed ${\rm AdS/CFT}$ along with appropriate twisted boundary conditions.

\end{minipage}

\end{center}

%%%%%%%%%%%%%%%%%%%%%%%%%%%%%%%%%%%%%%%%%%%%%%%%%%%%%%%%%%%%%%%%%%%%%%%%%%%%%%%%
\newpage
\setcounter{page}{1}
\renewcommand{\thefootnote}{\arabic{footnote}}
\setcounter{footnote}{0}

%\tableofcontents
%%%%%%%%%%%%%%%%%%%%%%%%%%%%%%%%%%%%%%%%%%%%%%%%%%%%%%%%%%%%%%%%%%%%%%%%%%%%%%%%

%%%%%%%%%%%%%%%%%%%%%%%%%%%%%%%%%%%%%%%%%%%%%%%%%%%%%%%%%%%%%%%%%%%%%%%%%%%%%%%%
%%%%%%%%%%%%%%%%%%%%%%%%%%%%%%%%%%%%%%%%%%%%%%%%%%%%%%%%%%%%%%%%%%%%%%%%%%%%%%%%
%%%%%%%%%%%%%%%%%%%%%%%%%%%%%%%%%%%%%%%%%%%%%%%%%%%%%%%%%%%%%%%%%%%%%%%%%%%%%%%%%%%%%%%%%
%                                                                                       %
%              Introduction ............                                                %
%                                                                                       %
%%%%%%%%%%%%%%%%%%%%%%%%%%%%%%%%%%%%%%%%%%%%%%%%%%%%%%%%%%%%%%%%%%%%%%%%%%%%%%%%%%%%%%%%%

\section{Introduction\label{sec1}}
The $S$-matrix plays a key role for studying two-dimensional integrable models.
With enough symmetries, the $S$-matrix can be determined mathematically and can be
used to find particle spectrum along with exact dispersion relations and
to compute finite-size effects.
Based on this philosophy, there have been remarkable developments in applying
the integrability methods to the AdS/CFT duality between
${\cal N}=4$ super Yang-Mills theory (SYM) and the type-IIB superstring
theory on ${\rm AdS}_5\times{\rm S}^5$ \cite{Beisert:2010jr}.
Exact $S$-matrix has been proposed \cite{Staudacher, sma, AFZ} with the dressing phase \cite{Janik, BES},
and applied to such tools as L\"uscher correction \cite{lusch} and thermodynamic Bethe asnatz \cite{TBA}.

After these successes, it is natural to extend the utility of the integrable
methods to other proposed or conjectured ${\rm AdS} / {\rm CFT}$ dualities.
These include $\beta$-deformed SYM theory \cite{LS95} which is dual to superstring theory on Lunin-Maldacena background \cite{LM05}
and three-parameter-deformed theory which breaks all the supersymmetry
\cite{F05}.
There are some clues that the deformations still maintain the integrability.
First, string sigma models on the deformed backgrounds are
classically integrable \cite{FrRoTs, FrRoTs1}.
One-loop dilatation operator on the gauge theory can be mapped to
integrable spin chain models with certain twists \cite{BerChe}.
All-loop asymptotic Bethe ansatz equations for the deformed theories were
conjectured by Beisert and Roiban \cite{BR05}.

Another strong evidence for the integrability has come from
the anomalous dimension of Konishi operator computed by
twisted L\"uscher formula \cite{Ahn:2010yv} which matches with four-loop perturbative computation \cite{FSSZ}.
Related computations have been also worked out by the Y-system
of the $\beta$-deformed SYM \cite{Gromov:2010dy}.

With the assumption of integrability, the $S$-matrix and associated twisted
boundary conditions have been proposed and used to derive the conjectured
all-loop asymptotic Bethe ansatz equations \cite{Ahn:2010ws}.
The twisted $S$-matrix is given by
\begin{eqnarray}
&&\tilde {\cal S}(p_{1},p_{2}) = F\cdot {\cal S}(p_{1},p_{2})\cdot F, \label{DrinfeldSmatrix}\\
&&F= e^{i \gamma_{1} \Gamma},\quad  \Gamma=h \otimes \id \otimes \id \otimes h -
\id \otimes h \otimes h \otimes \id ,\quad h=\diag(\frac{1}{2},-\frac{1}{2},0,0)\,, \nonumber
\end{eqnarray}
and the corresponding twisted boundary conditions are
\begin{eqnarray}
M_A = e^{i (\gamma_{3}-\gamma_{2}) J h_{A}}
\otimes e^{i (\gamma_{3}+\gamma_{2}) J h_{A}},
\label{DrinfeldBC}
\end{eqnarray}
where $A$ denotes an auxiliary space.

Another support of the $S$-matrix conjecture comes from the strong coupling
limit of the twisted AdS/CFT duality.
Finite $J$ correction of a classical giant magnon
dispersion relation has been computed from the $S$-matrix element and
twisted boundary conditions through L\"uscher formula \cite{Ahn:2012hs}
and shown to match
with classical sigma model computation for the $\gamma$-deformed
background \cite{Bykov:2008bj, Ahn:2011dq}.
While these evidences justify the assumption of integrability, it is
desirable to check the $S$-matrix directly
either with the string theory on a deformed background in the strong coupling limit
or with the ${\cal N}=1$ supersymmetric or non-supersymmetric gauge theories
in the weak coupling limit.

On the other hand, $S$-matrix in two dimensions which describes
localized interaction of two on-shell particle states and the boundary conditions imposed on the
particle states which characterize global spatial geometry are inseparable.
One can always attribute a part of $S$-matrix into the boundary conditions and
vice versa.
Consistency of this ambiguity is guaranteed only at the level of Bethe-Yang
equations which determine the energy level spectra of a theory.
In the context of the twisted AdS/CFT, it is possible to shift the
Drinfeld-Reshetikhin twist $F$ into the boundary condition $M$ as
shown in \cite{Ahn:2010ws}.
The resulting theory is described by untwisted $S$-matrix while the twisted
boundary condition is given by
\begin{eqnarray}
M_{A,Q_1,Q_2,\ldots,Q_N} =M_A\prod_{j=1}^{N}F_{AQ_j}^2.
\end{eqnarray}
This is called ``operatorial'' boundary condition since it depends not only the
particle state which passes through the boundary but also all other states $Q$'s in
the ``quantum space'' which are away from the boundary.
This feature is inevitable when one deals with off-diagonal $S$-matrix.
It is shown that this combination of $S$-matrix and boundary
conditions can produce the same ``Beisert-Roiban'' asymptotic Bethe ansatz
equations \cite{Arutyunov:2010gu}.

However, these operatorial boundary conditions are difficult to realize in the perturbative computations.
On string side, Frolov first showed that superstring theory on the $TsT$-transformed $AdS_5\times S^{5}$ with the periodic boundary conditions is equivalent to the undeformed $AdS_5\times S^{5}$ with the following twisted boundary conditions \cite{F05}:
\begin{equation}
\tilde{{\tilde \phi}}_i(2\pi)-\tilde{{\tilde \phi}}_i(0)=2 \pi (n_i + \epsilon_{ijk} \gamma_j J_k),\,\,(i,j,k=1,2,3).\label{TBC a}
\end{equation}
Here, $\gamma_{j}=\beta$ is a parameter for deformation of scalar field potential in the gauge theory side and three angular momenta are given by $J_i=\int dx {\dot \phi}_i$.
Eq.(\ref{TBC a}) is not easy to solve for the multiparticle solutions.
For the spin-chain side in the weak coupling limit, the $S$-matrix can be computed by
coordinate Bethe ansatz method.
The spin-chain Hamiltonian as a dilatation operator naturally depends on the deformation parameter $\beta$
in ${\cal N}=1$ SYM.
By some nontrivial unitary transformation it can be changed into that of untwisted spin-chain
as explained in \cite{BerChe}.
However, it generates nontrivial boundary conditions which will be in general nonlocal, i.e. which
depends on the states on the quantum space.

For these reasons, we study the $S$-matrix of the $\beta$-deformed SYM
at strong and weak 't Hooft coupling regimes which corresponds to (\ref{DrinfeldSmatrix})
where the boundary condition (\ref{DrinfeldBC}) becomes simply a $c$-number.
For this purpose, we consider string world-sheet action in near BMN limit and with light-cone gauge fixing
which is different from Lunin-Maldacena and compute the worldsheet scattering as was done
for untwisted case in \cite{KMRZ}.
In the weak coupling regime, we consider one-loop dilatation operator for three-spin sector.
We apply coordinate Bethe ansatz to compute one-loop $S$-matrix in this sector using the
deformed $SU(3)$ spin chain Hamiltonian derived in \cite{BerChe, Freyhult:2005ws}
and show that it matches with the exact Drinfeld-Reshetikhin $S$-matrix (\ref{DrinfeldSmatrix})
in this limit.

%%%%%%%%%%%%%%%%%%%%%%%%%%%%%%%%%%%%%%%%%%%%%%%%%%%%%%%%%%%%%%%%%%%%%%%%%%%%%%%%%%%%%%%%%
%                                                                                       %
%    Strong coupling regime : String worldsheet                                         %
%                                                                                       %
%%%%%%%%%%%%%%%%%%%%%%%%%%%%%%%%%%%%%%%%%%%%%%%%%%%%%%%%%%%%%%%%%%%%%%%%%%%%%%%%%%%%%%%%%

\section{Strong coupling regime : String worldsheet\label{sec2}}
The dual gravity solution of ${\cal N}=1$ $\beta$-deformed SYM is first constructed by Lunin and Maldacena \cite{LM05}. This background could be obtained by using sequence of three $TsT$-transformations : $(\phi_{1}, \phi_{2})_{TsT}$, $(\phi_{2}, \phi_{3})_{TsT}$ and $(\phi_{3}, \phi_{1})_{TsT}$ for three tori angles with single parameter ${\hat\gamma}=\gamma\sqrt{\lambda}, \,\ (\gamma_{i}=\gamma)$. Here, $(\phi_{1}, \phi_{2})_{TsT}$ means to take T-dualization along $\phi_{1}$, shift $\phi_{2}\rightarrow \phi_{2}+{\hat\gamma} \phi_{1}$ and take T-dualization again for $\phi_{1}$.
As a result of $TsT$-transformation, all kinds of background fields - metric, B-fields, RR-fields and so on - are deformed or generated.
If we use different parameters ${\hat\gamma}_{1,2,3}$ for each $TsT$-transformation, LM background could be generalized to three-parameter deformed background which is dual to non-supersymmetric, marginal deformed SYM.
The three-parameter deformed $AdS_5\times S^{5}$ spacetime metric and antisymmetric B-fields are given by the followings:
\begin{eqnarray}
ds^{2}_{{\rm string}}/R^{2} &=& ds^{2}_{AdS_{5}} + \sum_{i=1}^{3}\left(d\rho^{2}_{i}+G\rho^{2}_{i}d\phi^{2}_{i}\right)+G \rho^{2}_{1}\rho^{2}_{2}\rho^{2}_{3}\left(\sum_{i=1}^{3}\hat{\gamma}_{i} d\phi_{i}\right)^2, \cr
B_{2} &=& R^{2}G \left(\hat{\gamma}_{3}\rho_1^2 \rho_2^2 d\phi_1 \wedge d\phi_2
		+ \hat{\gamma}_{1}\rho_2^2 \rho_3^2 d\phi_2 \wedge d\phi_3
		+ \hat{\gamma}_{2}\rho_3^2 \rho_1^2 d\phi_3 \wedge d\phi_1\right)\,,\nonumber
\end{eqnarray}
where
\begin{equation}
	G^{-1} = 1 +  \hat{\gamma}_{3}^{2}\rho_1^2 \rho_2^2 + \hat{\gamma}_{1}^{2}\rho_2^2 \rho_3^2 + \hat{\gamma}_{2}^{2}\rho_3^2 \rho_1^2\, .\nonumber
\end{equation}
There is an additional constraint $\sum_{i=1}^{3} \rho_{i}^{2}=1$ and three tori angles $\phi_{1,2,3}$ have periodicity under $\sigma \rightarrow \sigma + 2\pi$.
We will only consider one-parameter deformed theory (${\hat \gamma}_{1,2,3}={\hat \gamma}$) for simplicity but all discussions about string regime in this paper are applicable even for three-parameter deformed theory.

\subsection{$TsT$-transformed $AdS_5\times S^{5}$ with twisted boundary conditions}
We start from $AdS_5\times S^{5}$ string with twisted boundary conditions (\ref{TBC a}).
The nonlinear sigma model action on usual $S^5$ is given by
\begin{equation}
\tilde{\tilde{S}} = -\frac{\sqrt{\lambda}}{2} \int d\tau \frac{d\sigma}{2 \pi} [h^{\alpha \beta}(\partial_{\alpha }\rho_{i}\partial_{\beta }\rho_{i}+\rho_{i}^{2} \partial_{\alpha } \tilde{\tilde{\phi}}_{i}\partial_{\beta }\tilde{\tilde{\phi}}_{i}+ \Lambda(\rho_{i}^{2}-1)].\label{ads5}
\end{equation}
Taking a $TsT$-transformation $(\tilde{\tilde{\phi}}_{2}, \tilde{\tilde{\phi}}_{3})_{TsT}$, we obtain a new background
\begin{eqnarray}
ds^{2}_{\rm string}/R^{2} &=& ds^{2}_{AdS_{5}} +d\rho^{2}_{1}+ \rho^{2}_{1}d{\hat \phi}^{2}_{1} +\sum_{i=2}^{3}\left(d\rho^{2}_{i}+{\hat G}\rho^{2}_{i}d{\hat\phi}^{2}_{i}\right),\cr
B_{2} &=& -\hat{\gamma}R^{2} {\hat G} \left(\rho_2^2 \rho_3^2 d{\hat\phi}_2 \wedge d{\hat\phi}_3\right), \quad {\hat G}^{-1} = 1 + \hat{\gamma}^{2} \rho_2^2 \rho_3^2 ,\label{M5}
\end{eqnarray}
with background metric $\hat{G}_{ij}$ and fields $\hat{B}_{ij}$ whose non-zero components are
\begin{equation}
\hat{G}_{11} = \rho_{1}^2 ,\quad  \hat{G}_{22} = {\hat G}\rho_{2}^2 ,\quad    \hat{G}_{33} = {\hat G}\rho_{3}^2 ,\quad  \hat{B}_{23} = {\hat G}\hat{\gamma}^2\rho_{2}^2\rho_{3}^2. \nonumber
\end{equation}
Corresponding bosonic string action is
\begin{eqnarray}
{\hat S} = - \frac{\sqrt{\lambda}}{2} \int d\tau \frac{d\sigma}{2 \pi}&& \left[h^{\alpha \beta}(\partial_{\alpha }\rho_{i}\partial_{\beta }\rho_{i}+\rho_{1}^{2} \partial_{\alpha }{\hat \phi}_{1}\partial_{\beta }{\hat \phi}_{1}+{\hat G}\rho_{2}^{2} \partial_{\alpha }{\hat \phi}_{2}\partial_{\beta }{\hat \phi}_{2}+{\hat G}\rho_{3}^{2} \partial_{\alpha }{\hat \phi}_{3}\partial_{\beta }{\hat \phi}_{3})\right.\cr
&&-\left. \epsilon^{\alpha \beta} \hat{B}_{{\hat \phi}_{2}{\hat \phi}_{3}}\partial_{\alpha }{\hat \phi}_{2}\partial_{\beta }{\hat \phi}_{3} + \Lambda(\rho_{i}^{2}-1)\right].\label{daction}
\end{eqnarray}
We will use this action to compute worldsheet $S$-matrix.\footnote{We restrict ourselves to bosonic fields only for simplicity.}

The above $TsT$-transformation changes the twisted boundary conditions (\ref{TBC a}) to
\begin{eqnarray}
\hat{\phi}_1(2\pi)-\hat{\phi}_1(0)&=& P_{ws}, \cr
\hat{\phi}_2(2\pi)-\hat{\phi}_2(0)&=& 2 \pi (n_{2}+\beta J_1),\cr
\hat{\phi}_3(2\pi)-\hat{\phi}_3(0)&=& 2 \pi (n_{3}-\beta J_1),\label{TBCb}
\end{eqnarray}
where level matching condition is given by $P_{ws}=2 \pi \left[n_{2}+\beta (J_3-J_2)\right]$.
This corresponds to ``c-number'' boundary conditions for $\hat{\phi}_{2}$ and $\hat{\phi}_{3}$ because they do not depend on  $J_{2}$ and $J_{3}$.

\subsection{Gauge fixed Lagrangian}
To compute the string worldsheet $S$-matrix, it is convenient to introduce new variables defined by
\be
\rho_1&=&\frac{\sqrt{{{Y}_1}^2+{{Y}_2}^2}}{1+\frac{Y^2}{4}},\,\,\,\,\,\, \rho_2=\frac{\sqrt{{Y_3}^2+{Y_4}^2}}{1+\frac{Y^2}{4}},\,\,\,\,\,\, \rho_3=\frac{1-\frac{Y^2}{4}}{1+\frac{Y^2}{4}},\cr
{\hat \phi}_2&=&\arctan({{Y}_2/{Y}_1}),\,\,\,\,\,\, {\hat \phi}_3=\arctan({{Y}_4/{Y}_3}),\,\,\,\,\,\, {\hat \phi}_1=\phi.
\ee
We also have to remove the redundancy from general coordinate invariance. A standard way is to consider the BMN limit \cite{BMN02} and its curvature corrections:
\begin{equation}
t \rightarrow X^{+}-\frac{X^{-}}{2 R^{2}},\,\,\,\,\, \phi \rightarrow X^{+}+\frac{X^{-}}{2 R^{2}},\,\,\,\,\, Z_{k} \rightarrow \frac{Z_{k}}{R},\,\,\,\,\, Y_{k'} \rightarrow \frac{Y_{k'}}{R}.
\end{equation}
This BMN limit simplifies the metric and B-fields as follows:
\begin{eqnarray}
ds^{2}&=&2dX^{+}dX^{-}+dY^2+dZ^2-{dX^{+}}^2(Z^{2}+Y^{2})\cr
&+&\frac{1}{2R^{2}}\left(2{dX}^{-}{dX}^{+}({Z}^{2}-{Y}^{2})+{dZ}^{2}Z^{2} - {dY}^{2}Y^{2}+{dX}^{+}{dX}^{+}(Y^{4}-Z^{4})\right),\cr
B&=& \frac{1}{2R^{2}}\hat{\gamma} (Y_{1}Y_{3}{dY}_{2}\wedge{dY}_{4}+Y_{2}Y_{4}{dY}_{1}\wedge{dY}_{3}
-Y_{2}Y_{3}{dY}_{1}\wedge{dY}_{4}-Y_{1}Y_{4}{dY}_{2}\wedge{dY}_{3}).\nonumber
\end{eqnarray}
Although the metric is independent of ${\hat \gamma}$ up to $1/R^2$, worldsheet scattering becomes nontrivial because the B-fields have $\hat{\gamma}$ dependence.\footnote{The same computation for original Lunin-Maldacena background shows that $(1+\hat{\gamma}^2)$ appears in front of $dY^2$ which gives different masses between ${AdS_5}$ and $S^5$ in gauge fixed action. This is one reason why we need to introduce twisted boundary conditions in string theory side.}

The bosonic string Lagrangian becomes now
\begin{eqnarray}
L&=&\frac{g}{2} \gamma^{\alpha \beta} \left[G_{++}\partial_{\alpha}X^{+}\partial_{\beta}X^{+} + G_{--}\partial_{\alpha}X^{-}\partial_{\beta}X^{-}\right.\nonumber\\
&+&\left. G_{+-}(\partial_{\alpha}X^{+}\partial_{\beta}X^{-}+\partial_{\alpha}X^{-}\partial_{\beta}X^{+}) + G_{Z^{i}Z^{j}}\partial_{\alpha}Z^{i}\partial_{\beta}Z^{j} + G_{Y^{i}Y^{j}}\partial_{\alpha}Y^{i}\partial_{\beta}Y^{j}\right]\cr
&+&  \frac{g}{2} \varepsilon^{ab} \left(B_{ij}\partial_{a}Y^{i}\partial_{b}Y^{j}+B_{+i}\partial_{a}X^{+}\partial_{b}Y^{i}+B_{-i}\partial_{a}X^{-}\partial_{b}Y^{i}\right),
\end{eqnarray}
with
\begin{equation}
\varepsilon^{ab}= \left(
\begin{array}{cc}
0 & 1 \\
-1 & 0  \\
\end{array}\right).\nonumber
\end{equation}
Here, $a$, $b$ stand for worldsheet coordinates $\sigma$ and $\tau$.

As in usual case, the Hamiltonian is just sum of Lagrange multiplier times constraint.
As the epsilon coupled to anti-symmetric $B$-fields is a non-dynamical field, the variation of the action over worldsheet metric operate on only $G$-field parts.

To fix the gauge, we can use the first-order formalism which works well for undeformed theory \cite{Callan:2003xr, Callan:2004uv, sundin}. First, we define the conjugate momentum
\begin{equation}
P_{\mu}=(\gamma^{\tau\sigma} G_{\mu\nu}+ B_{\mu\nu})\acute{X}^{\nu} + \gamma^{\tau\tau} G_{\mu\nu}\dot{X}^{\nu},
\end{equation}
and the Hamiltonian
\begin{equation}
H= \frac{1}{2 \gamma^{\tau\tau}} G^{\mu\nu} \bar{P}_{\mu}\bar{P}_{\nu} +\frac{1}{2 \gamma^{\tau\tau}} G_{\mu\nu} \acute{X}^{\mu}\acute{X}^{\nu} - \frac{\gamma^{\tau\sigma}}{\gamma^{\tau\tau}}\bar{P}_{\mu}\acute{X}^{\mu}, \,\ \bar{P}_{\mu}=P_{\mu}-B_{\mu\nu}\acute{X}^{\nu},
\end{equation}
which becomes zero if we impose the Virasoro constraints.
Introducing the light-cone gauge $X^{+}=\tau, \,\,\,\ P_{-}=const$, we can express the Lagrangian $L=P_{\mu} \dot{X^{\mu}}-H$
in terms of ungauged variables
\beq
L_{g. f.}=P_{+}+P_{I} \dot{X^{I}} = P_{I} \dot{X^{I}} - H_{L.C.}
\eeq
where we have imposed the Virasoro constraints.
The expression for the light-cone Hamiltonian is given by
\be
H_{L.C.}= {\tilde H}
&+&\frac{\hat{\gamma}}{\sqrt{\lambda}}
\left(-\acute{Y_{1}}Y_{2}Y_{3}P_{Y_4}+\acute{Y_{2}}Y_{1}Y_{3}P_{Y_4}
  +\acute{Y_{1}}Y_{2}Y_{4}P_{Y_3}-\acute{Y_{2}}Y_{1}Y_{4}P_{Y_3}
  +\acute{Y_{3}}Y_{4}Y_{1}P_{Y_2}\right.\cr
  &-&\left.\acute{Y_{3}}Y_{4}Y_{2}P_{Y_1}
  -\acute{Y_{4}}Y_{3}Y_{1}P_{Y_2}+\acute{Y_{4}}Y_{3}Y_{2}P_{Y_1}\right).
\ee
Here,  ${\tilde H}$ is the light-cone Hamiltonian of the undeformed theory.
Considering Legendre transformation and solving equations of motion for $P_{I}$,
we finally obtain the gauge fixed bosonic Lagrangian
\begin{equation}
L_{g.f.}= \frac{1}{2}{\partial_{\mu}{{\vec Z}^{\dag}}}\cdot{\partial^{\mu}{{\vec Z}}}  - \frac{1}{2} {{\vec Z}^{\dag}}\cdot{\vec Z}+
\frac{1}{2}{\partial_{\mu}{{\vec Y}^{\dag}}}\cdot{\partial^{\mu}{{\vec Y}}}  - \frac{1}{2} {{\vec Y}^{\dag}}\cdot{\vec Y} - {\mathbb V}({\vec Y},{\vec Z}), \label{gfact}
\end{equation}
with ${\vec Y}=(Y_{1\dot{1}},Y_{1\dot{2}},Y_{2\dot{1}},Y_{2\dot{2}})$ defined by
\be
Y_{1\dot{1}}&=&Y_{1}+ i Y_{2}, \quad Y_{1\dot{2}}=Y_{3}+ i Y_{4},\cr
Y_{2\dot{1}}&=&Y_{3}- i Y_{4}, \quad Y_{2\dot{2}}=Y_{1}- i Y_{2}.\nonumber
\ee
The potential term is
\be
&&{\mathbb V}=\frac{1}{4\sqrt{\lambda}}\left[(Y_{1\dot{1}}Y_{2\dot{2}}+Y_{1\dot{2}}Y_{2\dot{1}})(2{\partial_{\sigma}Y_{1\dot{1}}}{\partial_{\sigma}Y_{2\dot{2}}}
+2{\partial_{\sigma}Y_{1\dot{2}}}{\partial_{\sigma}Y_{2\dot{1}}}+(\partial_{\tau}Z)^2+(\partial_{\sigma}Z)^2)\right.\cr
&&-\left.Z^2({\partial_{\tau}Y_{1\dot{2}}}{\partial_{\tau}Y_{2\dot{1}}}+{\partial_{\tau}Y_{1\dot{1}}}{\partial_{\tau}Y_{2\dot{2}}}
+{\partial_{\sigma}Y_{1\dot{2}}}{\partial_{\sigma}Y_{2\dot{1}}}
+{\partial_{\sigma}Y_{1\dot{1}}}{\partial_{\sigma}Y_{2\dot{2}}}+2(\partial_{\sigma}Z)^2)\right]\cr
&&-\frac{\hat{\gamma}}{4\sqrt{\lambda}}\left[ Y_{1\dot{1}}Y_{1\dot{2}}{\partial_{\tau}Y_{2\dot{2}}}{\partial_{\sigma}Y_{2\dot{1}}}
-Y_{1\dot{1}}Y_{1\dot{2}}{\partial_{\tau}Y_{2\dot{1}}}{\partial_{\sigma}Y_{2\dot{2}}}
+Y_{1\dot{1}}Y_{2\dot{1}}{\partial_{\tau}Y_{1\dot{2}}}{\partial_{\sigma}Y_{2\dot{2}}}
-Y_{1\dot{1}}Y_{2\dot{1}}{\partial_{\tau}Y_{2\dot{2}}}{\partial_{\sigma}Y_{1\dot{2}}}\right.\cr
&&\left.+Y_{1\dot{2}}Y_{2\dot{2}}{\partial_{\tau}Y_{2\dot{1}}}{\partial_{\sigma}Y_{1\dot{1}}}
-Y_{1\dot{2}}Y_{2\dot{2}}{\partial_{\tau}Y_{1\dot{1}}}{\partial_{\sigma}Y_{2\dot{1}}}
+Y_{2\dot{1}}Y_{2\dot{2}}{\partial_{\tau}Y_{1\dot{1}}}{\partial_{\sigma}Y_{1\dot{2}}}
-Y_{2\dot{1}}Y_{2\dot{2}}{\partial_{\tau}Y_{1\dot{2}}}{\partial_{\sigma}Y_{1\dot{1}}}\right].\nonumber
\ee
To compute worldsheet $S$-matrix, we need to consider decompactification limit $P_{-}\rightarrow\infty$ in which worldsheet parameter space changes from cylinder to plane after rescaling $\sigma \rightarrow \frac{P_{-}}{\sqrt{\lambda}} \sigma$. Here, $P_{-}$ has appeared in the integration bound for $\sigma$ due to light-cone gauge fixing.

%%%%%%%%%%%%%%%%%%%%%%%%%%%%%%%%%%%%%%%%%%%%%%%%%%%%%%%%%%%%%%%%%%%%%%%%%%%%%%%%%%%%%%%%%
%                                                                                       %
%                          Tree Level Scattering Amplitudes from gauge fixed action     %
%                                                                                       %
%%%%%%%%%%%%%%%%%%%%%%%%%%%%%%%%%%%%%%%%%%%%%%%%%%%%%%%%%%%%%%%%%%%%%%%%%%%%%%%%%%%%%%%%%

\subsection{Tree-level scattering amplitudes\label{sec3}}
The string worldsheet $S$-matrix can be straightforwardly computed from the gauge fixed action (\ref{gfact}).
In the leading order of $\frac{1}{\sqrt{\lambda}}$, we define ${\mathbb T}$-matrix by
\begin{equation}
{\mathbb S}=\mathbb{I}+ \frac{2i \pi}{\sqrt{\lambda}} \mathbb{T}.
\end{equation}
We need to compute additional contribution to ${\mathbb T}$ from ${\hat \gamma}$-dependent part of ${\mathbb V}$ which contains only ${\vec Y}$.
In terms of mode expansions \cite{Arutyunov:2009ga}
\begin{eqnarray}
Y^{1\dot{1}}(\sigma,\tau) &=& \int \frac{dp}{2\sqrt{\omega_{p}}}\left[ a^{1\dot{1}}(p) e^{-i({\omega} \tau - p \sigma)}+\epsilon^{12}\epsilon^{\dot{1}\dot{2}}{a(p)}_{2\dot{2}}^{\dag} e^{i({\omega} \tau - p \sigma)}\right],\cr
Y^{1\dot{2}}(\sigma,\tau) &=& \int \frac{dp}{2\sqrt{\omega_{p}}}\left[ a^{1\dot{2}}(p) e^{-i({\omega} \tau - p \sigma)}+\epsilon^{12}\epsilon^{\dot{2}\dot{1}}{a(p)}_{2\dot{1}}^{\dag} e^{i({\omega} \tau - p \sigma)}\right],\cr
Y_{1\dot{1}}(\sigma,\tau) &=& \int \frac{dp}{2\sqrt{\omega_{p}}}\left[ \epsilon_{12}\epsilon_{\dot{1}\dot{2}}a^{2\dot{2}}(p) e^{-i({\omega} \tau - p \sigma)}+{a(p)}_{1\dot{1}}^{\dag} e^{i({\omega} \tau - p \sigma)}\right],\cr
Y_{1\dot{2}}(\sigma,\tau) &=& \int \frac{dp}{2\sqrt{\omega_{p}}}\left[ \epsilon_{12}\epsilon_{\dot{2}\dot{1}}a^{2\dot{1}}(p) e^{-i({\omega} \tau - p \sigma)}+{a(p)}_{1\dot{2}}^{\dag} e^{i({\omega} \tau - p \sigma)}\right],\nonumber
\end{eqnarray}
the ${\hat \gamma}$-dependent part of ${\mathbb T}$-matrix is
\be
{\mathbb T}_{\hat{\gamma}}&=& \hat{\gamma} \int \frac{dp dp'}{\Lambda(p,p')}
\left[(\omega p'-\omega' p){a(p)}_{1\dot{1}}^{\dag} {a(p')}_{2\dot{1}}^{\dag} {a(p)}_{1\dot{1}} {a(p')}_{2\dot{1}}
-(\omega p'-\omega' p){a(p)}_{1\dot{1}}^{\dag} {a(p')}_{1\dot{2}}^{\dag} {a(p)}_{1\dot{1}} {a(p')}_{1\dot{2}}\right.\cr
&+&\left.(\omega p'-\omega' p){a(p)}_{2\dot{2}}^{\dag} {a(p')}_{1\dot{2}}^{\dag} {a(p)}_{2\dot{2}} {a(p')}_{1\dot{2}}-
(\omega p'-\omega' p){a(p)}_{2\dot{2}}^{\dag} {a(p')}_{2\dot{1}}^{\dag} {a(p)}_{2\dot{2}} {a(p')}_{2\dot{1}}\right].\label{tmatrix}
\ee
Here, $\omega=\sqrt{p^2 + 1}$ and the kinematic factor \cite{Klose:2006dd}
\beq
\Lambda(p,p')= \frac{1}{\omega' p - \omega p'}.
\eeq

One can notice that the scattering amplitudes depend on $\hat{\gamma}$ only in $YY$ to $YY$ process.
Explicitly, only non-zero elements of the ${\mathbb T}_{\hat{\gamma}}$ are
\be
{\mathbb T}_{\hat{\gamma}} |Y_{1\dot{1}}(p)Y_{1\dot{2}}(p')\rangle = -\hat{\gamma} |Y_{1\dot{1}}(p)Y_{1\dot{2}}(p') \rangle, \cr
{\mathbb T}_{\hat{\gamma}} |Y_{1\dot{1}}(p)Y_{2\dot{1}}(p')\rangle = +\hat{\gamma} |Y_{1\dot{1}}(p)Y_{2\dot{1}}(p') \rangle, \cr
{\mathbb T}_{\hat{\gamma}} |Y_{2\dot{2}}(p)Y_{1\dot{2}}(p')\rangle = +\hat{\gamma} |Y_{2\dot{2}}(p)Y_{1\dot{2}}(p') \rangle, \cr
{\mathbb T}_{\hat{\gamma}} |Y_{2\dot{2}}(p)Y_{2\dot{1}}(p')\rangle = -\hat{\gamma} |Y_{2\dot{2}}(p)Y_{2\dot{1}}(p') \rangle. \label{nzt}
\ee

Now, we consider the strong coupling limit of the exact twisted S-matrix to compare with the above tree-level amplitudes.
In this limit, we can expand the twisted matrix $F$ for small $\beta=\hat{\gamma}/\sqrt{\lambda}$ \footnote{Originally, Lunin-Maldacena background was defined for small $\beta$.}
\begin{equation}
F = e^{\frac{{2 \pi i \hat{\gamma}}}{\sqrt{\lambda}}\Gamma} \simeq (\mathbb{I} + 2 \pi i \hat{\gamma}\frac{\Gamma}{\sqrt{\lambda}}),
\end{equation}
with $\Gamma$ defined in (\ref{DrinfeldSmatrix}) as well as the twisted $S$-matrix
\begin{equation}
\tilde {\cal S}= \mathbb{I} + 2\pi i \frac{(2 \Gamma \hat{\gamma} + \mathbb{T})}{\sqrt{\lambda}}, \label{stsm}
\end{equation}
where ${\mathbb T}$ is the undeformed matrix elements.
Because the elements of the twisted $S$-matrix (\ref{DrinfeldSmatrix}) can be written as
\begin{equation}
{\tilde {\cal S}_{ij}} = F_{il}\, {\cal S}_{lk}\, F_{kj} = F_{i} {\delta}_{il}\, {\cal S}_{lk}\, F_{k} {\delta}_{kj}=F_{i}\, F_{j}\, {\cal S}_{ij}, \nonumber
\end{equation}
only amplitudes which are deformed in two-boson to two-boson scatterings are
\begin{eqnarray}
{\tilde S}_{(1{\dot 1})(1{\dot 2})}^{(1{\dot 1})(1{\dot 2})}&=& e^{- \gamma}S_{11}^{11}\, S_{\dot{1}\dot{2}}^{\dot{1}\dot{2}}, \,\
{\tilde S}_{(1{\dot 1})(2{\dot 1})}^{(1{\dot 1})(2{\dot 1})}= e^{+ \gamma}S_{12}^{12}\, S_{\dot{1}\dot{1}}^{\dot{1}\dot{1}},\cr
{\tilde S}_{(2{\dot 2})(1{\dot 2})}^{(2{\dot 2})(1{\dot 2})}&=& e^{+ \gamma}S_{21}^{21}\, S_{\dot{2}\dot{2}}^{\dot{2}\dot{2}}, \,\
{\tilde S}_{(2{\dot 2})(2{\dot 1})}^{(2{\dot 2})(2{\dot 1})}= e^{- \gamma}S_{22}^{22}\, S_{\dot{2}\dot{1}}^{\dot{2}\dot{1}}.\nonumber
\end{eqnarray}
This matches with (\ref{nzt}).

On the other hand, we can get the twisted boundary conditions for ${\vec Y}$ from (\ref{TBCb})
\begin{eqnarray}
Y_{1\dot{1}}(+\pi P_{-}/\sqrt{\lambda})/Y_{1\dot{1}}(-\pi P_{-}/\sqrt{\lambda}) &=& e^{2 \pi i \beta J_1},\cr
Y_{1\dot{2}}(+\pi P_{-}/\sqrt{\lambda})/Y_{1\dot{2}}(-\pi P_{-}/\sqrt{\lambda}) &=& e^{-2 \pi i \beta J_1},\cr
Y_{2\dot{2}}(+\pi P_{-}/\sqrt{\lambda})/Y_{1\dot{1}}(-\pi P_{-}/\sqrt{\lambda}) &=& e^{2 \pi i \beta J_1},\cr
Y_{2\dot{1}}(+\pi P_{-}/\sqrt{\lambda})/Y_{1\dot{2}}(-\pi P_{-}/\sqrt{\lambda}) &=& e^{-2 \pi i \beta J_1},\nonumber
\end{eqnarray}
which also agree with (\ref{DrinfeldBC}) with $\gamma_3=\gamma_2=\beta$ and $J_1 =J$.

%%%%%%%%%%%%%%%%%%%%%%%%%%%%%%%%%%%%%%%%%%%%%%%%%%%%%%%%%%%%%%%%%%%%%%%%%%%%%%%%%%%%%%%%%
%                                                                                       %
%              Weak coupling regime : Spin-chains                                       %
%                                                                                       %
%%%%%%%%%%%%%%%%%%%%%%%%%%%%%%%%%%%%%%%%%%%%%%%%%%%%%%%%%%%%%%%%%%%%%%%%%%%%%%%%%%%%%%%%%

\section{Weak coupling regime : Spin-chains\label{sec4}}
The spin-chain Hamiltonian corresponding to the one-loop dilatation operator of the $\beta$-deformed SYM was first studied in \cite{BerChe, Roiban:2003dw}. Later, more general integrable deformation was investigated in \cite{Freyhult:2005ws}.
In this section, we compute $S$-matrix from the spin-chain Hamiltonian using coordinate Bethe ansatz. For simplicity, we only consider three-state spin-chain which is the simplest sector with nontrivial dependence on the deformation parameter.

The one-loop dilatation operator for the three-state operators $Z$, $X$ and $Y$ is given by \cite{BerChe}:
\begin{eqnarray}
H &=& \sum_{i=1}^{L} \left[-e^{2 \pi i  \beta}(E_{01}^{i}E_{10}^{i+1}+E_{12}^{i}E_{21}^{i+1}+E_{20}^{i}E_{02}^{i+1})
-e^{-2\pi i  \beta}(E_{10}^{i}E_{01}^{i+1}+E_{21}^{i}E_{12}^{i+1}+E_{02}^{i}E_{20}^{i+1})\right.\cr
&&\left.+(E_{00}^{i}E_{11}^{i+1}+E_{11}^{i}E_{22}^{i+1}+E_{22}^{i}E_{00}^{i+1})+(E_{11}^{i}E_{00}^{i+1}+E_{22}^{i}E_{11}^{i+1}+E_{00}^{i}E_{22}^{i+1})\right].\label{dila}
\end{eqnarray}
Here, the indices $0,1,2$ stand for the $Z,X,Y$ fields, respectively, and the matrix $E_{ab}$ is defined by $E_{ab}|c\rangle = |a\rangle \delta_{bc}$.
This Hamiltonian is integrable because it can be obtained from Drinfeld-Reshetikhin deformation of $SU(3)$ $R$-matrix.
For our purpose, it is more convenient to introduce a position-dependent unitary transformation \cite{BerChe}:
\begin{eqnarray}
&&|n\rangle_{0}\rightarrow|n\rangle_{0}, \quad |n\rangle_{1}\rightarrow e^{2\pi i \beta n}|n\rangle_{1}, \quad |n\rangle_{2}\rightarrow e^{-2\pi i \beta n}|n\rangle_{2}\cr
{\rm with} \quad &&|n\rangle_{a}\equiv|\cdots 00 \stackrel{\stackrel{n}{\downarrow}}{a} 00\cdots\rangle , \quad (a=0,1,2), \label{pdt}
\end{eqnarray}
which leads to a new Hamiltonian
\begin{eqnarray}
H &=& \sum_{i=1}^{L} \left[-(E_{01}^{i}E_{10}^{i+1}+e^{6 \pi i  \beta}E_{12}^{i}E_{21}^{i+1}+E_{20}^{i}E_{02}^{i+1})
-(E_{10}^{i}E_{01}^{i+1}+e^{-6\pi i\beta}E_{21}^{i}E_{12}^{i+1}+E_{02}^{i}E_{20}^{i+1})\right.\cr
&&\left.+(E_{00}^{i}E_{11}^{i+1}+E_{11}^{i}E_{22}^{i+1}+E_{22}^{i}E_{00}^{i+1})+(E_{11}^{i}E_{00}^{i+1}+E_{22}^{i}E_{11}^{i+1}+E_{00}^{i}E_{22}^{i+1})\right],\label{defh}
\end{eqnarray}
along with the twisted boundary conditions
\begin{equation}
|L+1\rangle_{0}=|1\rangle_{0}, \quad |L+1\rangle_{1}=e^{-2\pi i \beta L}|1\rangle_{1}, \quad |L+1\rangle_{2}=e^{2\pi i \beta L}|1\rangle_{2}.\label{aaaaa}
\end{equation}

To apply the coordinate Bethe Ansatz, we define an one-particle state
\begin{equation}
|\Psi\rangle_{a} =\sum_{n=1}^{L} e^{i p n} |n\rangle_{a}, \,\ (a=1,2).\nonumber
\end{equation}
Acting the Hamiltonian (\ref{defh}) on $|\Psi\rangle$,
we get the dispersion relation
\begin{equation}
E(p)=4 \sin^{2}\frac{p}{2}. \nonumber
\end{equation}
Now we consider two-particle scattering amplitudes. One can find easily that the $S$-matrix between two particles of the same type is the same as $SU(2)$ case,
namely
\begin{equation}
s(p_1,p_2)=\frac{u_1 - u_2 + i}{u_1 - u_2 - i}, \,\ \quad  u_k = \frac{1}{2} \cot\frac{p_{k}}{2}. \nonumber
\end{equation}
For two-particle states of different types $(a\neq b)$, we define
\begin{eqnarray}
|\Psi\rangle &=&\sum_{1\leq n_1\leq n_2 \leq L} \{\Phi_{12}(n_1, n_2)|n_1,n_2\rangle_{12}+\Phi_{21}(n_1, n_2)|n_1,n_2\rangle_{21}\}, \cr
\Phi_{12}(n_1,n_2)&=& A_{12}(p_1,p_2) e^{i (p_1 n_1 + p_2 n_2)}+A_{12}(p_2,p_1) e^{i (p_2 n_1 + p_1 n_2)},\cr
\Phi_{21}(n_1,n_2)&=& A_{21}(p_1,p_2) e^{i (p_1 n_1 + p_2 n_2)}+A_{21}(p_2,p_1) e^{i (p_2 n_1 + p_1 n_2)},\nonumber
\end{eqnarray}
where
\begin{equation}
|n_1,n_2\rangle_{ab}=|\cdots 00 \stackrel{\stackrel{n_1}{\downarrow}}{a} 00 \cdots 00 \stackrel{\stackrel{n_2}{\downarrow}}{b} 00\cdots\rangle.
\end{equation}
In terms of these amplitudes, we can define the $S$-matrix by
\begin{equation}
\left(
\begin{array}{c}
 A_{12}(p_2,p_1) \\
 A_{21}(p_2,p_1)
\end{array}
\right)=\left(
\begin{array}{cc}
 {\tilde r}(p_2,p_1) & {\tilde t}(p_2,p_1) \\
 {\tilde {\tilde t}}(p_2,p_1) & {\tilde{\tilde r}}(p_2,p_1)
\end{array}
\right)\cdot\left(
\begin{array}{c}
 A_{12}(p_1,p_2) \\
 A_{21}(p_1,p_2)
\end{array}
\right).
\end{equation}

From the eigenvalue equation $H |\Psi\rangle=E|\Psi\rangle$, we obtain
\begin{eqnarray}
0&=&A_{12}(p_1,p_2) e^{i p_2}\left(1-e^{-i p_2}-e^{i p_1}\right) +A_{12}(p_2,p_1) e^{i p_1}\left(1-e^{-i p_1}-e^{i p_2}\right)\cr
&&+e^{-6\pi i \beta}{A}_{21}(p_1,p_2) e^{i p_2}+e^{-6\pi i \beta}A_{21}(p_2,p_1) e^{i p_1},\label{tr1}\\
0&=&A_{21}(p_1,p_2) e^{i p_2}\left(1-e^{-i p_2}-e^{i p_1}\right) +A_{21}(p_2,p_1) e^{i p_1}\left(1-e^{-i p_1}-e^{i p_2}\right)\cr
&&+e^{6\pi i \beta}{A}_{12}(p_1,p_2) e^{i p_2}+e^{6\pi i \beta}A_{12}(p_2,p_1) e^{i p_1},\label{tr2}\\
{\rm with} \quad  E&=& E(p_1)+E(p_2). \nonumber
\end{eqnarray}
From the above equations, we can determine transmission and reflection coefficients as below:
\begin{eqnarray}
{\tilde r}(p_1,p_2)&=&{\tilde {\tilde r}}(p_1,p_2)= \frac{i}{u_1 - u_2 - i}, \cr
{\tilde t}(p_1,p_2)&=& \frac{u_1 - u_2}{u_1 - u_2 - i} e^{6\pi i \beta}, \cr
{\tilde {\tilde t}}(p_1,p_2)&=& \frac{u_1 - u_2}{u_1 - u_2 - i}e^{-6\pi i \beta}, \nonumber
\end{eqnarray}
and the twisted S-matrix for deformed three-spin states is given by
\begin{equation}
{\tilde S}_{\rm spin}= \left(
\begin{array}{cccc}
s(p_1,p_2) & 0 & 0 & 0 \\
0 & {\tilde t}(p_1,p_2) & {\tilde r}(p_1,p_2) & 0 \\
0 & {\tilde {\tilde r}}(p_1,p_2) & {\tilde {\tilde t}}(p_1,p_2) & 0 \\
0 & 0 & 0 & s(p_1,p_2) \\
\end{array}\right).\label{tsm}
\end{equation}

The $S$-matrix (\ref{tsm}) agrees with the weak coupling limit of the exact $S$-matrix (\ref{DrinfeldSmatrix}) except ${\tilde t}, {\tilde {\tilde t}}$. This discrepancy can be attributed to the frame factors assigned differently for spin-chain and worldsheet scatterings which happens also for undeformed case \cite{AFZ, KMRZ}.
For the $\beta$-deformed case, the $S$-matrices for the $SU(3)$ sector are related by
\begin{equation}
{\tilde S}_{\rm string}=U(p_1)\cdot \,\, {\tilde S}_{\rm spin}\,\, \cdot U(p_2)^{-1}, \nonumber
\end{equation}
where the frame factor $U(p)$ is given by
\begin{equation}
U(p)=\left(
\begin{array}{cccc}
e^{i p}e^{2\pi \beta i} & 0 & 0 & 0\\
0 & e^{i p}e^{-2\pi \beta i} & 0 & 0\\
0 & 0 & e^{i p}e^{-2\pi \beta i} & 0\\
0 & 0 & 0 & e^{i p}e^{2\pi \beta i}\\
\end{array}
\right).\nonumber
\end{equation}
It is straightforward to check ${\tilde S}_{\rm string}$ agrees with $\lambda\rightarrow 0$ limit of (\ref{DrinfeldSmatrix}).

%%%%%%%%%%%%%%%%%%%%%%%%%%%%%%%%%%%%%%%%%%%%%%%%%%%%%%%%%%%%%%%%%%%%%%%%%%%%%%%%%%%%%%%%%
%                                                                                       %
%                           Conclusions and Discussions                                 %
%                                                                                       %
%%%%%%%%%%%%%%%%%%%%%%%%%%%%%%%%%%%%%%%%%%%%%%%%%%%%%%%%%%%%%%%%%%%%%%%%%%%%%%%%%%%%%%%%%

\section{Conclusions\label{sec5}}
In this paper we have computed worldsheet and spin-chain scatterings of the $\beta$-deformed SYM in the leading order to check the validity of proposed exact $S$-matrix and boundary conditions.
For the strong 't Hooft coupling regime, we used the light-cone gauge fixed Lagrangian in the $TsT$-transformed background.
We also computed weak coupling $S$-matrix based on $SU(3)$ spin-chain Hamiltonian.
We have shown that these perturbative results match with the exact conjectures.

Here, we have considered only boson to boson scatterings in the leading order. It will be interesting to extend the checks to fermions and the higher-loop order.
It will be also interesting to investigate whether our simpler background of the $\beta$-deformed theory can be more useful in finding concrete string solutions or higher correlation functions.

%%%%%%%%%%%%%%%%%%%%%%%%%%%%%%%%%%%%%%%%%%%%%%%%%%%%%%%%%%%%%%%%%%%%%%%%%%%%%%%%%%%%%%%%%
%                                                                                       %
%                                   Acknowledgements                                    %
%                                                                                       %
%%%%%%%%%%%%%%%%%%%%%%%%%%%%%%%%%%%%%%%%%%%%%%%%%%%%%%%%%%%%%%%%%%%%%%%%%%%%%%%%%%%%%%%%%

\section*{Acknowledgements}
We thank to Zoltan Bajnok, Diego Bombardelli, Rouven Frassek, Tristan Mcloughlin, Rafael Nepomechie and Per Sundin for valuable discussions and comments.
This work was supported in part by WCU Grant No. R32-2008-000-101300 (C. A.),
and by the National Research Foundation of Korea (NRF) grant funded by the Korea government (MEST)
through the Center for Quantum Spacetime (CQUeST) of Sogang University with grant number 2005-0049409
(M. K. and B.-H. L.).

\appendix

%%%%%%%%%%%%%%%%%%%%%%%%%%%%%%%%%%%%%%%%%%%%%%%%%%%%%%%%%%%%%%%%%%%%%%%%%%%%%%%%%%%%%%%%%
%                                                                                       %
%                                   Appendix : ABBN S-matrix and bdry condition       %
%                                                                                       %
%%%%%%%%%%%%%%%%%%%%%%%%%%%%%%%%%%%%%%%%%%%%%%%%%%%%%%%%%%%%%%%%%%%%%%%%%%%%%%%%%%%%%%%%%

%\bibliographystyle{nb}
%\bibliography{paper}

\end{document}